\documentclass[12pt,A4]{article}

\begin{document}

\vspace {1.0 in}

\title{ Physics and Complexity}
\author{David Sherrington \\
Rudolf Peierls Centre for Theoretical Physics \\
University of Oxford,
1 Keble Road, Oxford OX1 3NP, UK \\
and \\
Santa Fe Institute, 1399 Hyde Park Road, Santa Fe, NM 87501, USA }

\maketitle

\begin{abstract}

This paper is concerned with complex macroscopic behaviour arising in 
many-body systems through the combinations of competitive interactions 
and disorder, even with simple ingredients at the microscopic level. 
It attempts to indicate and illustrate the richness 
that has arisen, in
conceptual understanding, in methodology and in application, across 
a large range of scientific disciplines, together with a hint of 
some of the further opportunities that remain to be tapped. In 
doing so it takes the perspective of physics and tries to 
show, albeit rather briefly, how physics has contributed and 
been stimulated.

\end{abstract}

\section{Introduction}

Already for more than 50 years, particularly since the theoretical 
explanation of superconductivity by Bardeen, Cooper and Schrieffer 
\cite{BCS} and as highlighted by Philip Anderson  in his seminal 
essay ``More is different''\cite{Anderson72}, it has been recognised 
that the cooperative physics of interacting systems can exhibit 
fundamental new behaviour that is not apparent in the properties 
of the individual `elementary units'  that make up a many-body system 
of a very large number of these units; for example as found in 
condensed matter made up of atoms, ions and electrons.  Furthermore, 
important features of this new `emergent' behaviour are often 
independent of the details of the make-up, for example whether a 
system is fairly `pure' or rather imperfect; one remarkable example 
is the sharp quantization of plateaux in the Hall effect of low 
temperature slightly disordered semiconductors \cite{QHE}, providing 
the most accurate method of measuring the fundamental constant ${e^2}/h$. 

Here, however, I shall discuss the emergence of a different phenomenon, 
known as `complexity', concentrating on the new concepts 
that have arisen from its discovery and study and particularly 
the role of physics as a mindset 
and a methodology in its exploration and application. 

The story has its origin in an attempt to understand the behaviour of 
some relatively obscure magnetic alloys which have never themselves 
had any technological application but whose study has led to highly 
non-trivial insights of considerable subtlety, new methods of thinking, 
new mathematical, experimental and simulational techniques, and, in 
some cases, important applications, not only in many areas of physics 
itself but also in information processing, hard combinatorial optimization, 
biology, mathematics and economics, with much further and broader potential. 

As noted earlier, the systems of interest are `many-body', made 
up of many ($N >> 1$) similar individual units, with the concern 
the co-operative behaviour of the whole. The descriptor `complex' is used 
to describe collective behaviour that cannot be anticipated simply from the 
properties of isolated individual units or from interactions 
among only a few of them, but arises from conflicts when 
large numbers of individuals have mutually incompletely satisfiable 
few-body rules, a feature known as `frustration'; indeed complex
cooperative behaviour can arise with even very simple individual units
and very simple interactions. Among the 
consequences of this frustration and the resultant compromise 
are that optima and equilibrium are difficult to achieve and 
that responses to perturbations are slow, in part extremely so, 
and often chaotic.  

A `cartoon' to illustrate the character of a 
complex system is of a rugged landscape with many hills and valleys 
and the system's dynamics imagined as movement on the 
landscape with only local vision, simple moves being 
downward\footnote{Physicists usually think in terms of cost functions 
to minimise, while evolutionary biologists think in terms of 
complementary fitnesses to maximise. I shall use the physics 
convention; the mapping between them is just a minus sign.}, 
becoming stuck in intermediate valleys and unable 
to surmount ridges, needing a change of rule to overcome  
(for example, allowing also uphill moves), only to be faced 
by further barriers. In fact, the space of the landscape is 
very high and the difficulty extremely much greater than for 
a mountain-range on our three-dimensional Earth. Changing an 
external influence parameter, such as temperature, magnetic 
field or other `pressure' can lead to `chaotic' transformation 
of the whole landscape, removing 
the option of straightfoward iteration to a solution.

This paper is organized as follows; Section 2 introduces the issues 
and some of the concepts derived in the context of 
spin glasses (in condensed matter physics) and compares
 a related optimization problem
which is simple to formulate but difficult to solve; Section 3 considers 
further hard combinatorial optimization in computer science, 
particularly in the context of satisfiability; Section 4 is devoted 
to some examples from biology, in the form of neural networks and proteins;
Section 5 addresses problems involving speculative agents
in finance/economics  and opens the way to other social science; Section 6
considers growing networks; Section 7 gives some typical magnitudes;
Section 8 collects some conclusions and hints at the future.

\section{The Dean's Problem and Spin Glasses}

A simple illustration is provided by
the so-called `Dean's problem' in which a 
College Dean is faced with the task of placing students into two 
dormitories in such a way as to ensure that the students are as happy 
as possible, given that any individual pair of students might want 
to be in the same dorm or in different ones\footnote{For
simplicity we shall assume that feelings between pairs of students 
are mutual.}.
 If for any three students 
the number of pairing preferences for being apart is odd then not all 
these preferences can be satisfied simultaneously. This is `frustration' 
and there is no unique best choice. With a large number of students, 
finding the best compromise is very difficult, indeed in general 
NP-complete\footnote{
%Non-deterministic polynomially hard, with the 
%number of steps needed to solve effectively 
%growing exponentially in $N$. 
NP is the class of problems where we can verify, but not necessarily find, 
a solution in time polynomial in the number of elements $N$. 
NP-complete problems are the hardest ones in NP,
and are believed to take exponential time to solve.
Note that in the Dean's problem the total number of possible 
choices is $2^N$ and we are interested in large $N$.}
\cite{Clay2000, Garey}.

In fact, the Dean's problem was first considered in the context of 
the magnetic alloys mentioned earlier, as a minimalist model for a 
so-called `spin glass'\cite{Binder}. The original spin glasses are 
substitutional metallic alloys 
such as ${\rm Au}_{(1-x)}{\rm Fe}_{x}$, where only the Fe ions carry 
magnetic moments (or `spins') and, as a function of their separation, 
their pair-wise interactions are a mixture of ferromagnetic, trying 
to align the two spins, or anti-ferromagnetic, trying to make them 
point in opposite directions. Experimentally \cite{Mydosh} these 
materials were observed to exhibit a phase transition to an unusual 
state, with frozen moments but no periodic order - hence the 
appellation `glass' by analogy with amorphous window glass - , slow 
to respond to changes in external controls, accompanied by 
non-ergodicity, behaving differently depending on the order in 
which external perturbations, such as magnetic field or temperature, 
are applied. Nowadays slow response and non-ergodicity\footnote{An ergodic 
system equilibrates and its behaviour under the application of more than one 
perturbation is independent of the order in which those 
perturbations are applied. In a non-ergodic system there is no complete 
equilibration and the order of application matters; a standard example for
a spin glass is the combination of application of a magnetic field and of 
a change of temperature; beneath a critical `spin glass' temperature the 
susceptibility measured by first cooling and then applying a field 
(`field-cooled' - FC - susceptibility) is greater than that that obtained by the 
opposite order of events (zero field cooled - ZFC - susceptibility).}, 
along with memory, aging and rejuvenation effects,  
are considered the principal characteristics warranting the label `glass' 
and the expression `spin glass' is now used much more broadly to refer to 
systems which exhibit such glassiness due the combination of 
quenched disorder and frustration.

Both a minimal idealization of a spin glass and the Dean's problem 
can be modelled 
with a control function \cite{EA,SK}
\begin{equation}
H=-\sum_{(ij)}J_{ij}\sigma_{i}\sigma_{j}; \sigma =\pm 1
\label{eq:SK}
\end{equation}
where the $i,j$ label spins (students), ${\sigma=\pm 1}$ indicates 
spin up/down (dorm A/B),
and the interactions (preferences) $\{J_{ij}={J_{ji}}\}$ 
are chosen randomly and independently (i.i.d.) from a 
distribution $P_{exch}(J_{ij})$.
In the case of the Dean's 
problem
$H$ is a cost function to be minimized. In 
the case of the spin glass $H$ is the Hamiltonian energy function and 
in thermodynamic equilibrium at a temperature $T$ the probability of 
a microscopic state ${\{\sigma_i\}}$ is given by
\begin{equation}
P(\{\sigma\})=Z^{-1}\exp\{-H(\{\sigma\})/ T\}
\label{eq:P(s)}
\end{equation}
where Z is the partition function
\begin{equation}
Z=\sum_{\{\sigma\}} \exp\{-H(\{\sigma\})/ T\}.
\label{eq:Partition_function}
\end{equation}
At $T=0$ the problems coincide\footnote{More precisely, the Dean's 
problem corresponds to the SK spin glass where the sum in eqn (\ref{eq:SK}) 
is over all $(ij)$ and each $J_{ij} (=J_{ji})$ is drawn independently from the same 
distribution. To model a short-range spin glass the $\{i,j\}$ are chosen as 
neighbours on a lattice of appropriate dimensionality;  three for normal 
experience.}, while eqn (\ref{eq:P(s)}) can also be interpreted in a 
Dean's problem as the distribution resulting if the Dean allows a 
corresponding probabilistic degree of imprecision in 
his/her attempt to satisfy the students\footnote{In fact, introduction of an 
analogue of temperature has proven to be a very valuable tool in 
optimization, using so-called `simulated annealling' \cite{Kirkpatrick} 
in which stochastic imprecision is deliberately introduced into a computer 
updating algorithm in order to stochastically overcome barriers in an 
simple energy/cost description, 
and is then gradually reduced.}.

When the temperature $T$ and the mean $J$ are less than critical 
values (proportional to the variance of $J$) there results complex behaviour. 
A similar complexity is present in many other systems describable by 
a combination of competitive interaction and disorder, including 
situations where the disorder is effectively self-induced through 
`sticking' of the dynamics.

Having already noted that the Dean's problem is NP-complete one 
might reasonably ask how one knows much about the properties of these 
systems. One answer is through computer simulation, another is
through rather subtle mathematics and physical interpretation.
Also, an 
important point in these developments is that in physics one is normally 
interested in typical systems and their statistical 
properties, rather 
than `worst case', and the 
NP-hard label strictly applies to the worst case problem. One expects
the properties of a `good' large many-body system to depend 
on the statistical distribution from which its elements are drawn
rather than the specific instances that result. This does not 
however mean that the typical case problem is trivial - indeed 
the situation is very much the contrary, both technically and 
conceptually - and in fact the insight 
gained from typical case study has been very influential in explaining 
several intriguing empirical features found in several optimization 
problems and in predicting others, as well as in stimulating the 
development of new powerful techniques applicable even for 
specific (non-averaged) instances of hard problems. Typical behaviour 
study enables a powerful (and novel) set of tools to be applied and 
often allows the determination of the value of the best achievable 
average cost
without the need of an actual algorithm to achieve it in specific 
cases.  

The detailed theoretical methodologies devised to study spin 
glasses are beyond the scope of this article; the reader is 
referred instead to books such as \cite{MPV, Young, DG}. It is, 
however, appropriate to mention that they have involved the 
introduction of unusual (and, at least initially, non-rigorous) 
mathematics and ans\"atze, often 
guided by physical insight as well as driven pragmatically by a lack of 
alternatives to progress at the time. These irregular procedures led 
to predictions that regularly turned out to be confirmed 
in computer simulations of the models, as well as similar to 
experimental features, and provided valuable 
insights for new experiments with new types of probes. 
For the infinite-range spin glass\footnote{Also 
known as the mean field spin glass.} 
some of these results
have
recently been proven with rigorous mathematics while others are 
still being investigated by mathematicians\footnote{So far 
all the results obtained rigorously for infinite-range spin glasses 
\cite{Talagrand} have been in 
accord with 
the earlier less rigorously deduced results arising
from Parisi's ansatz\cite{Parisi}.}, 
although 
some subtle aspects remain controversial for spin glasses 
with short-range interactions (including  the  
experimental solid state systems that first inspired the 
studies). Perhaps most importantly, however, these  
studies have 
also led to wide conceptual insights and technical applications 
in an enormous range of complex systems. 

Despite the not inconsiderable mental and mathematical 
contortions that led to our current understanding, it is 
now possible to give physical explanation for the rugged landscape 
picture for the mean field spin glass model and its 
extensions. To this end we imagine that, at the temperature of 
interest, our system has a number of essentially separated 
macrostates - let us label them by an index $S$. In macrostate 
$S$ the thermal average of microscopic variable $\sigma_i$ is 
denoted by $\langle \sigma_i \rangle_{S}$, which in general 
can be either positive or negative and varies for different $i$
or $S$. 
One defines the `overlap' between two macrostates $S,S'$ as 
\begin{equation}
q_{SS'}=N^{-1}\sum_{i} \langle \sigma_i \rangle_{S} 
\langle \sigma_i \rangle_{S'}.
\label{eq:overlap}
\end{equation}
The overlap distribution is given by
\begin{equation}
P_{\{J_{ij}\}}(q)=\sum_{S,S'}W_S W_S' \delta (q-q_{SS'}),
\label{eq:overlap-distribution}
\end{equation}
where $W_S$ is the probability of macrostate $S$. 
In general the macrostates depend on the specific choice of the 
$\{J_{ij}\}$ but for the SK model both the averaged $P(q)$ and 
its higher moments can be calculated, as also other more 
complicated distributions of the $q_{SS'}$, such as the correlation 
of pairwise overlap distributions for 3 macrostates $S,S',S''$.
The 
first interesting observation is that beneath a critical `spin 
glass ordering' temperature the average $P(q)$ has structure,
with weight at more than one value of $|q|$, indicating that there 
are many non-equivalent  relevant macrostates, in contrast 
to the situation in normal single-phase systems, paramagnetic, 
ferromagnetic or antiferromagnetic, where 
$P(|q|)$ is a single 
delta-function. This structure is known also as `Replica Symmetry
Breaking' (RSB), following the mathematical method 
first employed to investigate it \cite{Parisi}. Secondly, although 
simple physical 
observables, like the energy, are self-averaging the overlap 
distribution is not. Thirdly, the distribution of overlaps involving 
three states shows that they exhibit the feature of `ultrametricity', 
the two smallest overlaps being equal and signalling a hierarchical 
state structure \cite{MPSTV}. These observations suggest a rugged landscape 
picture with barriers impenetrable on timescales becoming infinite 
with $N$. For finite-ranged spin glasses this picture must be relaxed  
to have only finite barriers, but still leading to 
slowness on physical timescales.

$P(q)$ cannot yet be measured directly in normal thermodynamic 
experiments but 
it can in computer simulations \cite{Young83}, 
by running and cross-comparing identical 
model systems evolving separately. Also, remarkably, it turns out to 
be indirectly measurable in dynamics away from equilibrium \cite{Parisi 2005},
as a 
correction to the normal fluctuation-dissipation relation. For 
range-free spin glasses one can also solve for other aspects of the 
dynamics away from equilibrium. This demonstrates that correlation 
and response functions become non-stationary in the spin glass phase 
and exhibit interesting `aging' and memory effects \cite{CuKu, Vincent}.  
Again these 
evocative `pictures' spread over and are observed also in other systems.

Before passing to other areas of application it is probably worth 
re-emphasising that the models leading to these understandings were 
{\it{minimalist}}, with as few parameters as possible; the simplest 
kind of `spins', just two-state; the simplest type of interactions, 
two-spin; 
the simplest relevant disorder characterization, independently 
chosen randomly from identical distributions characterised by 
the minimum number of 
parameters. This immediately demonstrates 
that the rich complexity found is a consequence of the 
collective behaviour and not due to any complication of 
properties of the individual units.
{\it{`Complex' is different from `complicated'!}} Immediately 
this suggests that in other potentially complex systems one 
can advantageously progress by looking for their minimal 
frustrated models, rather than including at the outset all aspects of 
real systems at the microscopic level.    

Many other systems in condensed matter physics are now 
recognised as having conceptually related glassy complexity\footnote{ 
Although the `spins' may be pseudo-descriptions of other physical 
attributes; see {\it{e.g.}} \cite{DS_2008,Ren}.}, 
but let us now turn to different subjects with different `rules'.

\section{Combinatorial optimization - satisfiability}

The topic of combinatorial optimization has already been introduced 
through the Dean's problem, but in fact there are a very great 
many hard optimization problems of considerable interest and intrigue.
One such class, much considered in computer science, 
is known as `satisfiability'. Here the task is to find
the values of  $N$ binary variables $x_{i};\, i=1..N; \, x=1,0$ such
that a set of constraint `clauses' is satisfied; {\it{e.g.}}
\begin{equation}
(x_{1}\;  \rm{or} \; {\bar{x}}_{2} \; \rm{or} \; x_{3}) \; \rm{and} 
\; ({\bar{x}}_{3} \; \rm{or} \; x_{4} \; \rm{or} \; {\bar{x}}_{5}) 
\; \rm{and} \; (......) \; \rm{and} \; .....
\label{eq:K-SAT}
\end{equation} 
where $x_{i}$ denotes $x_{i}=1$ (or true) and ${\bar{x}}_{i}$ denotes 
$x_{i}=0$ (or false).
Here the sections between round brackets are the `clauses' and the 
number of 
$x$-alternatives to be satisfied within a clause is usually denoted 
$K$ (so in the example
shown $K=3$) and the problem is referred to as K-SAT. It 
is a classic NP-hard problem for $K \ge 3$ \cite{KS}. The interest 
is particularly in
when the number of clauses $M$ scales as $M=\alpha N$ with $\alpha$ 
independent of $N$ and $N$ large. In random K-SAT the choices of 
the allocations of the $x_{i}$ and 
$\bar{x}_i$ to the clauses are random (and quenched). The typical 
case becomes 
unsatisfiable for $\alpha$ greater than a critical value 
$\alpha_c$. 

Random K-SAT corresponds to an extension of the SK model with 
spins interacting 
in a combination of 
terms of the form ${\sigma}_{i_1}...{\sigma}_{i_p}$ where 
$1 \le p \le K$ and the $i$ are chosen from $1...N$; this yields 
a graph structure which is a `cactus' extension 
of the random graphs  of Erd\"os and R\'enyi\cite{Erdos}\footnote{A 
further convenient graphical description employs factor 
graphs with two kinds of vertices corresponding respectively 
to the $x$ and to their interaction groupings \cite{Mezard-Montanari}.}. 
Already, therefore, one can expect analogous glassy
effects in an attempt to solve the random K-SAT problem
using a computer algorithm based on local operations. 
In fact, however, the $p$-spin version of the SK model exhibits 
further interesting behaviour beyond the $p=2$ SK original.
In particular, there are several transitions as the temperature 
of this $p$-spin model is reduced. Two of these are thermodynamic;
the higher, at $T_{G}^{1}$, is to a state known as `one step 
replica symmetry broken' (1RSB), 
corresponding to the onset of many non-equivalent macrostates $\{S\}$ 
that are mutually orthogonal \cite{Parisi}, 
followed by a lower temperature transition, at $T_{G}^F$,
to `full replica symmetry breaking' (FRSB) \cite{Gardner}
at which the non-equivalent 
macrostates start to acquire a
continuous range of overlaps (as characterises 
the SK spin glass phase). These transitions are however pre-empted by
a dynamical transition at a higher temperature, $T_{D}$,
again to 1RSB. The fact that 
$T_{D} > T_{G}^{1}$ means that the system becomes glassy before the 
thermodynamic limit for paramagnetism is reached. 

The analogue in Random K-SAT is that, although 
there is an in principle
limit ${\alpha}_c$ separating satisfiable (SAT) from unsatisfiable (UNSAT),
in practice computer programs based on local (dynamical) 
algorithms are unable to 
reach this limit but dynamically stick at a lower $\alpha_D$, the 
analogue of $T_{D}$. The recognition of  the existence and 
quantification of the resultant intermediate HARD-SAT region 
came through the 
application and extension to K-SAT of statistical physics 
techniques developed for spin 
glassses \cite{Mezard-Montanari, MPZ}. In fact the technique,  
known as `survey propagation' (and a development of the original 
cavity method \cite{MPV}), can often be applied also to specific as 
well as averaged 
problems and has proven extremely useful practically\footnote{It is 
useful when  
small loops in the graph stucture are 
not important.}. 

Furthermore, there are other 
quasi-transitions in the $p$-spin glass, for example at a 
temperature higher than $T_{D}$ 
a $p$-spin glass acquires an extensive configurational entropy which
affects response and correlation functions without being 
thermodynamically relevant,  again with an  analogue in K-SAT \cite{Krzakala}.
Studies of the $p$-spin glass in an applied magnetic field demonstrate that 
beyond a critical field the dynamical and thermodynamical transitions
come together. One might therefore usefully consider the possibility of 
including an analogue of such a field in a computer algorithm to assist
in the solution of hard optimization problems with effective $K \ge 3$,
removing HARD-SAT. Complementarily, computer scientists categorize 
several types of NP and it is natural to look for reflections in physics.

Finally in this section, one might note that the $p-$spin 
situation, with the first thermodynamic transition to 1RSB, appears to 
be the norm in many extensions to other spin glass situations, 
particularly those 
characterized by a lack of `symmetry' between 
`ferromagnetic' and `antiferromagnetic' interactions, such as with Potts
or quadrupolar spins, beyond critical spin dimensionality. It is also
believed to characterize the behaviour 
self-induced disorder in structural glasses.

\section{Neural networks and proteins}

Multi-valley landscapes are not always a nuisance - in fact they can 
be very valuable.
A good example is found in Hopfield's simple model \cite{Hopfield} 
for biological neural networks where 
the valleys provide storage for memories and where dynamics within the 
valleys represents memory 
retrieval. In this model one imagines the `brain' as made up of a very large 
number of neurons, firing to different degrees, 
connected by a large number of `synapses', both excitatory and inhibitory, 
the former such that a firing efferent neuron tends to make the afferent neuron fire 
and the latter such that a firing efferent neuron tends to 
reduce the firing of the afferent neuron.  If the synapses are symmetric 
the collective dynamics of the neurons can  
be envisaged as dominantly downward motion on a high-dimensional 
landscape whose structure is 
determined by the collection of synapses. Many valleys allow for 
many memories, a clear requisite 
for a useful brain, and many valleys requires frustration. That 
these valleys are in the high-dimensional many-neuron `space' 
provides `extended memory' and consequent robustness against 
failure of individual 
neurons or synapses. 

In the spirit of eqn.(\ref{eq:SK}) Hopfield 
idealized to $N$
binary (McCulloch-Pitts) neurons (firing/non-firing; $\sigma_{i}=\pm1$) 
and, in the spirit of Hebb \cite{Hebb}, took the synapses to be
related to stored patterns through
${\xi^{\mu}_{i};\mu=1,..P={\alpha}N}$ 
by
\begin{equation}
J_{ij}=P^{-1}\sum_{\mu}\xi^{\mu}_{i} \xi^{\mu}_{j}
\label{eq:Hopfield}
\end{equation}
with the neural state distribution given by eqn (\ref{eq:P(s)}) where the 
`temperature' $T$ characterises a sigmoidal rounding of the 
response of an afferent
neuron's firing to the sum of its inputs  from all its many feeding efferent 
neurons, via their corresponding synapses \cite{Amit}. The 
valleys then correspond to the memories $\{\sigma_i =\xi^{\mu}_i\}$ and the 
process of retrieval is the 
approach towards the minima of these valleys. Hence there arises 
the concept of `attractors', transcending the idealization of the modelling. 
Two
control parameters act to separate retrieval and non-retrieval phases, the 
`temperature'
$T$ and the `capacity' $\alpha$. The former can be viewed as 
`synaptic noise', the 
latter as `interference noise' arising from the frustrating effects 
of competing memories.
More generally, the identification of eqn (\ref{eq:Hopfield}) is 
not required, as neither 
is the symmetry $J_{ij}=J_{ji}$, although without this symmetry 
the landscape picture 
of a minimizable Lyapunov function is no 
longer strictly true and a more subtle attractor picture is needed. 

Thus far we have imagined the interactions as `quenched', fixed in time. For 
most practical purposes this is true for the experimental solid state 
spin glasses. 
However, for a neural network it is clear that there should be learning 
as well 
as retrieval, but on a rather slower timescale. This can be achieved 
conceptually by 
permitting the synapses to evolve in response to stimuli. Indeed, 
since the structures of 
 the cartoon landscapes (including the locations of valleys) 
 are determined by the 
synapses, it follows
that such synaptic response to external stimuli is essential to learn 
for future recognition 
and generalization - {\it{retrieval is motion on a landscape, 
learning is the sculpture of the landscape}}. 

Other examples of systems with some frustration and with both 
fast and slow dynamics can  
be found in other aspects of biological behaviour. For example, 
proteins are heteropolymers 
with competing interactions between their amino acids 
and both hydrophilic and 
hydrophobic elements. These competitions give rise to frustration. 
Consequently
random heteropolymers are, in general, poor folders 
(in analogy with the slow 
dynamics and sticking of spin glasses), in contrast to 
proteins which must fold quickly. It has been
suggested \cite{Wolynes} that proteins are a special class of 
heteropolymers with `minimal frustration', often described 
in cartoon terms by a `folding 
funnel' with secondary valleys along its sides (due to 
the remanent frustration). One can speculate 
that the proteins arose through a random 
process of evolution in which, from an initial `soup' of random 
hetero-polymers, as well as subsequent mutations, the 
successful folders are selected through being better able to fold
and hence reproduce. Similarly it seems likely that often survival depends
on unprogrammed but self-selected mutual 
`assistance'\footnote{For example, autocatalytic sets \cite{Kauffman_2}}.

\section{Econophysics}

More interesting examples of complex systems are found in 
economic and financial systems, the topic of the new science 
of `econophysics'. In these systems the `units' are people 
(or groups of people or institutions), speculators, 
producers and consumers, competing and cooperating, having different
individual  
strategies and inclinations ({\it e.g.} of degrees of risk-aversion or 
social conscience) and effectively interacting through information,
much of it commonly available (news, internet, stock-prices, 
etc.). The full economic system is complicated, but an interesting 
minimalist model inspired by stockmarkets is provided by the `minority 
game'(MG) \cite{Challet}. In this model $N$ `agents' play a game 
in which at each step each agent makes one of two choices 
(`buy' or `sell') with the aim to make the minority choice 
over all agents\footnote{The philosophy is to emulate the 
idea that one 
gets the best price by selling when most people want to buy or 
buying when most people want to sell.}. They make their choices based on a  
a commonly available piece of `information' but acted upon 
using personal 
strategies. In the original version the
information is the sequence 
of minority choices over the previous $m$ time-steps and each agent  
has two strategies, Boolean operators which acting on the information 
string yield an action-decision, and the choice of which strategy to 
employ is determined through a personal point-tally which, at each 
time-step, 
rewards strategies that give the actual minority choice; in a 
deterministic version each agent plays his/her strategy with 
the highest point-tally at the time. 
The minority requirement corresponds 
to frustration, while disorder arises through quenched i.i.d. random 
choices of strategies allocated to the agents.  
Simulational studies of the `volatility', the standard deviation 
of the number buying minus the number selling 
at each timestep, show that  over many time-steps agents are 
effectively correlated with one another (since it differs 
from the value that would result 
from agents making independent random buy/sell choices). Rather, 
the volatility 
has a cusp-minimum at a critical value of the ratio of the 
information dimension to the number of agents $\alpha = D/N = 2^{m}/N$, 
suggesting a phase transition at $\alpha_c$. Furthermore, the
transition is 
ergodic for $\alpha > \alpha_c$, non-ergodic for $\alpha < \alpha_c$\footnote{As
characterised by the irrelevance or relevance of a bias in the 
initial point-scores.}.  Another interesting observation is 
that the behaviour is almost identical if the true history of the last $m$ 
timesteps is replaced by a purely fictitious `random' history provided that 
each agent acts as though that fictitious history were true and additionally
that it spans the same information-space. {\it{Collective behaviour 
is determined by universal belief rather than truth!}} 

The minority game can be studied by methods developed in the theory of 
spin glasses, although it should be hastened to add that there are differences.
First, averaging the point-score dynamics over timescales greater than order 
$N$ enables the elimination of the explicit information to 
yield a temporally-coarse-grained point-dynamics involving 
a control function of a similar 
structure as that of eqn (\ref{eq:SK}), but also including a random field,  
\begin{equation}
H_H=-\sum_{(ij)}J_{ij}\sigma_{i}\sigma_{j} - 
\sum_{i}h_{i}\sigma_{i}; 
\label{eq:MG}
\end{equation}
where both the $J_{ij}$ and the $h_{i}$ are 
related to the randomly chosen strategies. In a slightly different 
formulation of the MG the strategies are expressible as points
on the corners of a $D$-dimensional hypercube 
${\xi_{i}^{\mu}; \mu =1,..D;\xi =\pm 1}$. In this formulation
\begin{equation}
J_{ij}=-D^{-1}{\sum_{\mu =1}^{D}} \xi_{i}^{\mu} \xi_{j}^{\mu}.
\label{MG_J}
\end{equation}
This is reminiscent of the form of the 
Hopfield model (\ref{eq:Hopfield}). However, 
there is an important difference in the sign, making the {$\{\xi\}$} of 
eqn. (\ref{eq:Hopfield}) repellors rather than attractors.
    
$H_H$ can be minimised using methods devised for spin glass thermodynamics,
but for more general behaviour, both ergodic and non-ergodic,
 it is 
necessary to use techniques developed for spin glass macrodynamics. 
At least part of this procedure is possible using a generating 
functional method \cite{Coolen}, yielding for large $N$ 
an effective single agent 
stochastic ensemble behaviour     
\begin{equation}
p(t+1) = p(t) -\alpha \sum_{t' \leq t} ({\bf{1}} + {\bf{G}})^{-1}_{tt'} 
\mbox{sgn}p(t') + {\sqrt \alpha} \eta(t),
\label{eq:effectiveagent}
\end{equation}
where $\eta(t)$ is coloured noise determined self-consistently over the 
corresponding ensemble by
\begin{equation}
\langle \eta(t) \eta(t') \rangle = [({\bf{1}} + 
{\bf{G}})^{-1}({\bf{1}}+{\bf{C}})({\bf{1}}+{\bf{G^T}})^{-1}]_{tt'}
\label{eq:colourednoise}
\end{equation}
and ${\bf{G}}$ and ${\bf{C}}$ are ensemble-self-consistently determined
response and correlation functions. Note that this ensemble has both 
memory and stochastic noise whereas the original many-body problem had 
neither and also that this mapping is valid for both
ergodic and non-ergodic regimes \cite{GS}\footnote{The many-body 
problem could be generalised to include memory and stochasticity
but the point here is that they are not necessary to produce them in
the effective single-agent ensemble.}. {\it{The concept of a single 
deterministic/rational effective agent is false, but that of an effective agent 
ensemble is true!}} 

The real world of finance and economics is significantly more 
complicated than the simplistic Minority Game. However, (i) already one sees 
several conceptually important aspects that are likely to extend to 
reality; one is the inadequacy of the 
doctrine of `rational 
expectations' in which it is assumed that given the same information
every agent will make the same deduction and achieve 
equilibrium - it has already been pointed out that even for the MG there is no
representative agent but only a representative stochastic ensemble - and 
another 
is clear from the recognition that even if agents could change their 
strategies there is no optimal solution in which 
all have the same strategy, since, by definition, not everyone 
can be in the minority, (ii) the model could be extended in increments 
of `reality' or complication, for example including the accumulation and 
expenditure/gambling of varying degrees of 
capital, the effects of market impact, 
allowing for producers, who must both buy and sell in order to carry out their
business, and  for consumers\footnote{Often referred to as `liquidity providers',
the financial analogue of energy input, such as provided by ATP in a 
biological context.}, or involving limit-order-book bidding, 
settling and price 
movement  (iii) the inclusion of 
some of these effects will 
result in there never being any `equilibrium', although there might be 
periods of relative stochastic quasi-steady-state, albeit probably
also punctuated by catastrophic crises, (iv) many of the 
drivers of cooperation involve information with no spatial-range 
restriction, thereby 
justifying mean-field theory treatment as a good approximation, 
(v) one 
could investigate the introduction of global `friction' and `bias'
mechanisms, as for example through the imposition of trading taxes and/or 
other political and legal constraints and incentives. 

Indeed one can imagine many other social science problems involving 
many individuals, often grouped into intermediate local,
political or national groups, where the interest is in collective behaviour 
under the influence of both heterogeneous inherent/cultural inclinations and 
national and international
laws. Clearly, anticipated reactions to variation of the laws is of crucial 
importance in trying to anticipate the resultant `health' of the whole. Minimal 
models and their incremental complication play an important 
role in separating the essential from the peripheral, in anticipating 
the likelihood of catrastrophes and trying to devise `dampers' to 
avoid or minimise them. 
Disorder, in the guise of different inherited and culturally-nurtured 
inclinations such as risk-aversion, honesty and
emotionality, as well as mental and physical abilities, is part 
of the natural make-up of individuals, while frustration is a feature of their 
collective co-existence requiring a degree of compromise to better the whole. 

 \section{Scale-free networks}
 
 Thus far, the discussion has assumed that the actual microscopic units, 
 as well as
 their connectivity and/or locations, do not change. Within this restriction 
 there have been considered several structures; crystalline/lattice 
 (canonical experimental
 spin glass alloys and the Edwards-Anderson model), fully connected 
 (Sherrington-Kirkpatrick, p-spin and Hopfield models) and 
 Erd\"os-R\'enyi graph structures (K-SAT). Simple extensions within the
 same general classes would include amorphous solid structures and random 
 graphs with i.i.d. probabilistic occupation of bonds \cite{Viana-Bray, 
 Fu-Anderson, Banavar}, as well as Low Density Parity Check (LDPC) 
 error-correcting codes \cite{Gallagher, MacKay}\footnote{For 
 error-correcting codes a theoretical limit on the necessary redundancy 
 to be able to correct errror was already discovered long ago by 
 Shannon \cite{Shannon} but has proven impossible to reach practically 
 - the reason is that the correction problem is frustrated, as well
 as effectively disordered, and the Shannon limit is the analogue of
 a thermodynamical transition while the algorithms to try to 
 reach it practically are dynamical with consequent glassy hindrance.}. 
 But 
 the last decade has seen an emergence of significant interest in 
 another type of structure, known as scale-free  \cite{Calderelli, Newman} 
 and epitomised by many 
 social and biological networks which have often grown spontaneously in 
 response to need - canonical examples include the internet, 
 hub-based airline systems, protein-protein interaction networks. 
 In these normally-random networks the distribution of node connectivities 
 behaves as $P(k) \approx k^{-\gamma}$ where $k$ is the degree of connectivity
 (number of other nodes connected directly to the one under consideration)
 and $P(k)$ is its statistical distribution. A further extension 
 concerns `networks under churn' in which nodes both enter and leave 
 dynamically - an example is found in peer-to-peer computer
 networks. In these networks,
 varying the relative rates of entry and leaving can lead to topological
 changes of global structure from scale-free to exponential \cite{Bauke}. Study of consequences of frustration and
 disorder in interactions of entities located on 
 such networks is relatively in its infancy.

\section{Magnitudes}

As emphasised earlier, our interest is in systems of a large 
number of individuals. 
Some models are analytically soluble, at least in principle or in part, in
limits
where the number of  
individuals $N$ is very large, $N \to \infty$, the 
effective number of 
interactions $M=\alpha N$  is also large and the interactions are
range-free. These restrictions are rarely strictly true but study
of such systems  
does provide
 conceptual and some quantitative insight into real systems. 
 It is perhaps 
 worth noting some typical actual sizes. In solids $N$
 is typically of the order of Avogadro's number ($O(10^{23})$) while
 $\alpha$ is of order $10$ and the range is usually finite; in the 
 brain there are of the order $10^{10}$ neurons, $\alpha$ is of order $10^5$
 and the synaptic range is (relatively) long; in economics/finance 
 the number of 
 interacting agents or trading units can be large, the number of effective 
 interactions enormous\footnote{There are typically no direct 
 interactions but, as noted earlier, individual 
 interaction with a high-dimensional and varying 
 information field provides an effective inter-agent interaction with 
 $\alpha=N$.} and the range infinite, but effects of 
 finiteness of $N$ can be relevant\footnote{{\it e.g.} in the MG 
 increasing the number of players while maintaining the information 
 dimension can lead to a transition through the critical $\alpha$ separating
 ergodic and non-ergodic regimes, thereby providing a natural limit 
 on the usefulness of entering the fray.}; proteins vary in length from a 
 few hundred amino acids ({\it{e.g.}} haemoglobin A has 287, yeasts 
 have of order 400 - 500) to order tens
 of thousands ({\it{e.g.}} titins in muscle). 
 
\section{Conclusions}

In this brief perspective I have attempted to give an 
impression of the conceptual,
mathematical, experimental and simulational challenges and 
novel discoveries that 
the combination of 
disorder and frustration
in many body systems has yielded, together with hints of 
some of the application opportunities their recognition 
has offered and continues to offer. The scientific journey of discovery 
has been highly 
interdisciplinary and there is much more scope, in many directions.  

This paper has emphasised the 
start from and perspective
of physics, but there have been other starts (and are surely
other perspectives) in other subjects;
for example, in
evolutionary biology the early work of 
Kauffman on genetic regulatory networks \cite{Kauffman},
employing random Boolean networks; 
the important work of Shannon on the achievable limit in 
error-correcting code theory 
has already been mentioned (in a footnote)
but not his seminal development of the whole subject of information theory;
computer scientists  
developed concepts like NP-hardness,  a plethora of theorems, 
bounds and 
equivalences, and many methodologies including 
belief propagation - equivalent to mean field theory. However, 
it is only in the last couple of decades that the opportunities provided 
by the  combination of
perspectives and expertises across the disciplines 
have been properly realised and started to be harnessed
and driven. 

These  developments have relied also on a mutually supportive and 
productive interplay of three different modes of study. Two of these 
were previously  established as methodologies, namely mathematical theory
and experimentation on real systems, although the types of 
 studies discussed above have led to 
many new approaches within these areas. But it has also seen the birth of 
and helped to drive as an equally important partner in the quest
another mode of investigation that was previously little represented. This 
is computer simulation as an investigatory tool on idealised models, 
able to make observations 
not possible in real experiments; this is to be contrasted with the use of 
computers to simulate complicated `realistic models' to emulate 
real experiments and with numerical analysis of theoretical expressions 
or experimental data, both of which have of course grown
with the corresponding increase in technical
computing power. For example, 
one can effectively perform experiments on systems whose microscopic 
properties are known exactly, typically corresponding to the same 
minimalist models as studied
theoretically, without the complications of real nature and 
with the possibility to vary from real nature ({\it{e.g.}}
different microscopic dynamics) and to employ probes 
(measurement methods) for which no analogue currently exists in 
the `real' laboratory but whose knowledge could check and/or guide theory.
Complementarily, the need for such simplifications of 
`quasi-experimental' systems and 
unphysical probes to understand complex systems has led to simulational
techniques which might not otherwise have been developed but 
which have proven very valuable.

There remain many further opportunities for both scientific 
understanding and practical application. Some can already be anticipated,
but others are yet to be thought of; the developments of 
the last 3 decades have led to so many 
remarkable and unanticipated discoveries that it seems inevitable 
that many more will arise. The next 
few decades offer the 
prospect of much richness for scientific explorer 
and technological applier alike.

%\section*{References}
%

%
\end{document}